\documentclass[twocolumn,showpacs,preprintnumbers,prl,amsmath,amssymb]{revtex4}

\usepackage{graphicx}
\usepackage{bm}

\newcommand{\Sr}{SrFe$_2$As$_2$}

\newcommand{\tc}{$T_c$}
\newcommand{\hc}{$H_{c2}$}

\newcommand{\etal}{{\it et al.}}

\begin{document}


\title{Superconducting and ferromagnetic phases induced by lattice distortions in SrFe$_2$As$_2$}


\author{S.~R.~Saha, N.~P.~Butch, K.~Kirshenbaum,}
\author{Johnpierre~Paglione}
\email{paglione@umd.edu}
\affiliation{Center for Nanophysics and Advanced Materials, Department of Physics, University of Maryland, College Park, MD 20742}
\author{P. Y. Zavalij}
\affiliation{Department of Chemistry and Biochemistry, University of Maryland, College Park, MD 20742}

\date{\today}


\begin{abstract}
Single crystals of SrFe$_2$As$_2$ grown using a self-flux solution method were characterized via x-ray, transport and magnetization studies, revealing a superconducting phase below $T_c = 21$~K characterized by a full electrical resistivity transition and partial diamagnetic screening. The reversible destruction and reinstatement of this phase by heat treatment and mechanical deformation studies, along with single-crystal X-ray diffraction measurements, indicate that internal crystallographic strain originating from $c$-axis-oriented planar defects plays a central role in promoting the appearance of superconductivity under ambient pressure conditions in $\sim90$\% of as-grown crystals. The appearance of a ferromagnetic moment with magnitude proportional to the tunable superconducting volume fraction suggests that these phenomena are both stabilized by lattice distortion.

\end{abstract}


\maketitle


The discovery of high-temperature superconductivity in Fe-based
pnictide compounds with ZrCuSiAs-type structure has sparked a flurry
of activity in the condensed matter physics community. The
suppression of magnetic/structural order in LaOFeAs by fluorine
substitution was shown to promote superconductivity below 26~K in
LaO$_{1-x}$F$_x$FeAs at ambient pressures \cite{Kamihara} which
increases to 43~K under applied pressures \cite{Takahashi}. The
highest transition temperature was subsequently increased even
further via rare earth substitutions, reaching 55~K in
SmO$_{1-x}$F$_x$FeAs \cite{Ren1}. More recently, a closely related
series of oxygen-free FeAs-based compounds with the
ThCr$_2$Si$_2$-type (122) structure have also exhibited similar
phenomena, with superconducting transition temperatures up to  $\sim
37$~K promoted by chemical substitution of alkaline or transition
metal ions \cite{Sasmal,Rotter,Jeevan,Sefat,Leithe,Saha} or by the
application of large pressures \cite{CaFe2As2,Alireza,Kumar}. To
date, ambient-pressure superconductivity has only appeared in a few
stoichiometric FeAs-based 122 compounds \cite{Rotter2,Sasmal} and
only below 4~K, suggesting that structural and/or chemical tuning is
vital to stabilizing high-temperature superconductivity in these
materials. It is widely believed that this is intricately tied to
the suppression of the magnetic/structural phase transition common
to all these materials, with structural tuning playing a key role in
determining the superconducting critical temperature of the
ferropnicitides  \cite{bondangle}.

Here we report the provocative observation of superconductivity below 21~K in self-flux-grown single crystals of the undoped parent compound \Sr\ at ambient pressures, deep below the tetragonal-to-orthorhombic structural phase transition and coexistent with antiferromagnetic order. Through systematic heat and pressure treatments and magnetization measurements, we find evidence of the controllably reversible appearance of this superconducting phase along with a small ferromagnetic moment that appears proportional to the superconducting volume fraction estimated from diamagnetic screening. Together with single-crystal X-ray crystallography, we uncover a surprising relationship between the appearance of these phenomena and the level of lattice distortion tied to preferentially oriented planar defects, suggesting an intimate relationship between superconductivity, magnetism and crystallographic strain in this system of materials.


Single-crystalline samples of \Sr\ were grown using the FeAs
self-flux method \cite{Wang}. The FeAs binary precursor was first
synthesized by solid-state reaction of Fe (5N) and As (4N) powders
in an evacuated quartz tube, then mixed with elemental Sr (3N5) in
the ratio 4 : 1 and heated in a quartz tube (either evacuated or
filled with partial atmospheric pressure of As) to 1100$^\circ$C,
grown both with and without the use of alumina crucibles. Chemical
analysis was obtained via both energy- and wavelength-dispersive
X-ray spectroscopy, showing 1:2:2 stoichiometry in all specimens
reported herein and no indication of impurity phases or any other
differences between superconducting (SC) and non-superconducting
(non-SC) samples. Structural properties were characterized by both
powder and single-crystal X-ray diffraction measured at 250~K with
Mo K-alpha radiation using Rietfeld (TOPAS-4) and single crystal
(SHELXS-97) refinement, yielding lattice constants
$a=3.9289(3)$~\AA\ and $c = 12.320(2)$~\AA\ with $R=1.36\%$ goodness
of fit. Reciprocal lattice precession images were generated from
Bruker APEX2 software. Basal-plane resistivity was measured with the
standard four-probe ac method using silver paint contacts made at
room temperature in all cases except for pressure experiments, where
silver epoxy contacts were made by curing at 200$^{\circ}$C for 5
minutes. Magnetic susceptibility $\chi$ was measured in a commercial
SQUID magnetometer.


\begin{figure}[!t]\centering
       \resizebox{8cm}{!}{
              \includegraphics{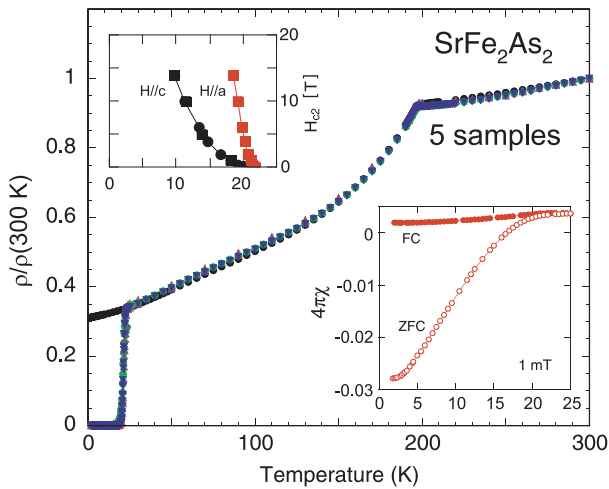}}
              \caption{\label{rho} Comparison of resistivity of several as-grown specimens
              of \Sr\ normalized to 300~K, showing a complete superconducting transition in four of five samples.
              Top inset: \hc$(T)$ deduced from extrapolated resistive onset of \tc\ in two samples for two field orientations.
              Bottom inset: magnetic susceptibility of superconducting sample measured with $H=1$~mT$\|$c.}
\end{figure}

Fig.~\ref{rho} presents the comparison of the resistivity $\rho(T)$
of several as-grown samples of \Sr\ normalized at room temperature
to remove geometric factor errors. As shown, all samples exhibit
identical metallic behavior with a sharp kink at $T_0=198$~K, which
has been associated with a structural phase transition (from
tetragonal to orthorhombic upon cooling) that is coincident with the
onset of antiferromagnetic (AFM) order \cite{Yan}. At lower
temperatures, a sharp drop in $\rho(T)$ beginning at 23~K is
consistently found in almost all as-grown samples synthesized under
varying heating schedules. While a minority of non-SC samples do not
show any trace of a resistive transition down to 18~mK, the majority
of our samples consistently show a transition at the same
temperature, with a midpoint at $T_c=21$~K and a large, anisotropic
upper critical field \hc\ as shown in the top inset of
Fig.~\ref{rho}.

Superconductivity in this undoped ``parent'' compound is surprising, given that charge doping or applied pressure are necessary to invoke similar transition temperatures throughout this family of materials \cite{Sasmal,Rotter,Jeevan,Sefat,Leithe,CaFe2As2,Alireza,Kumar}, and its intermittent appearance in different samples suggests that it is an unstable or parasitic phase, similar to the appearance of filamentary superconductivity at 1~K far above the bulk transition of 0.4~K in the heavy-fermion superconductor CeIrIn$_5$ \cite{Bianchi}.
However, etching and sanding the surfaces of our samples yields no difference in $\rho(T)$, and six-wire $\rho(T)$ measurements (configured to measure voltage drops due to currents forced to travel through the thickness of the sample) suggest that the superconductivity is not confined to the sample surfaces. More important, shown in the bottom inset of Fig.~\ref{rho}, the onset of a small but distinct diamagnetic response in magnetic susceptibility is consistent with the onset of partial  volume fraction superconductivity below 21~K. As stated below, we observe this fraction to reach as high as 15~\%, suggesting a non-negligible volume of bulk superconductivity can be stabilized in \Sr.


\begin{figure}[!t] \centering
  \resizebox{8cm}{!}{
  \includegraphics[width=8cm]{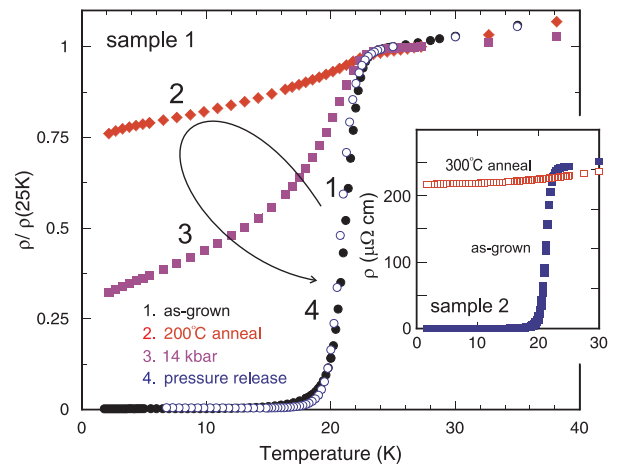}}
  \caption{\label{anneal} Heat treatment and applied pressure effects on resistive superconducting transition in one sample of \Sr, following the sequence: 1) unannealed, as-grown sample, 2) 200$^{\circ}$C heat treatment, 3) 14~kbar applied pressure, and 4) ambient conditions after release of pressure. Inset: a different superconducting sample subjected to 300$^{\circ}$C heat treatment.}
\end{figure}

This phase appears to be extremely sensitive to sample treatment: with moderate heat and pressure treatments, the resistive transition can be controllably removed and reinstated. Fig.~\ref{anneal} presents $\rho(T)$ of a superconducting sample subject to 200$^{\circ}$C treatment in air for only 5 minutes, which causes an almost complete suppression of the full superconducting drop originally observed in the as-grown sample. The inset of Fig.~\ref{anneal} shows the same effect on the resistivity of a second sample subject to 2 hours at 300$^{\circ}$C in flowing argon (using same contacts), which completely erases any trace of superconductivity. Interestingly, heat treatment affects only the magnitude of the resistance drop and not \tc\ itself, suggesting that it is not tuning the pairing amplitude but rather the volume fraction of superconductivity in the sample. This is corroborated by other independent observations of kinks in $\rho(T)$ positioned at exactly the same \tc\ in \Sr\ samples grown under quite different conditions, including polycrystalline \cite{Sasmal} and Sn flux-grown crystals \cite{Torikachvili2}, and recently, even in isoelectronic  BaFe$_2$As$_2$ \cite{Tanatar}.

Surprisingly, the suppression effect is reversible: upon application of external pressure, the suppressed resistive transition appears to return as evidenced by a stronger partial transition in $\rho(T)$, with subsequent recovery of a full transition upon release of applied pressure as shown in Fig.~\ref{anneal}. A second experiment that involved subjecting an annealed sample (originally superconducting) to severe ($\sim 100\%$) mechanical deformation invoked by a press, also showed evidence for a return of a partial transition in $\rho(T)$ upon subsequent measurement.


\begin{figure}[!t] \centering
  \resizebox{7cm}{!}{
  \includegraphics{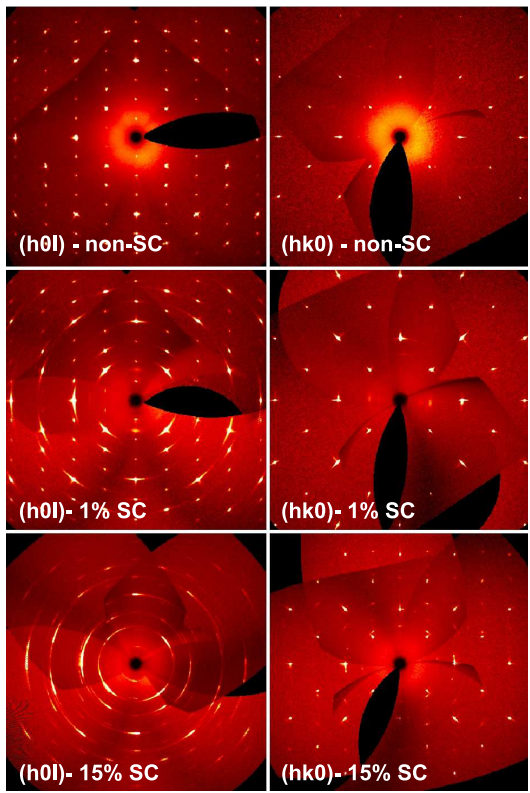}}
  \caption{\label{xray} Reciprocal lattice structure of \Sr\ from single-crystal x-ray diffraction measurements at 250~K for three characteristic samples with varying measured superconducting volume fractions as noted. Note that significant changes occur in the (h0l) reflection zone, but none are observable in the (hk0) c-axis projection. }
\end{figure}

Together, these data suggest a form of internal crystallographic strain as the likely origin of the superconducting instability in undoped, unpressurized \Sr, which is released upon annealing and reinstated upon cold-working. Defect structures have indeed been observed in other compounds with the ThCr$_2$Si$_2$ structure, such as URu$_2$Si$_2$ \cite{Ramirez}. To investigate this, we performed single-crystal X-ray diffraction measurements on three characteristic samples exhibiting 0, 1 and 15\% superconducting volume fractions.
As shown in Fig.~\ref{xray}, the reciprocal lattice data indeed show evidence for the appearance of crystal lattice distortions with increasing volume fraction, as evidenced by an increasing ring-like smearing of diffraction spots. Interestingly, this smearing is present only in the (h0l) patterns and not the (hk0) patterns, suggesting a form of planar defect that lies preferentially along the basal plane causing a distribution of orientations tilted about the basal plane axes. This scenario is corroborated by the recent report of enhanced $c$-axis resistivities in superconducting samples of \Sr\ and BaFe$_2$As$_2$ crystals \cite{Tanatar}.


\begin{figure}[!t] \centering
  \resizebox{8.5cm}{!}{
  \includegraphics[width=8.5cm]{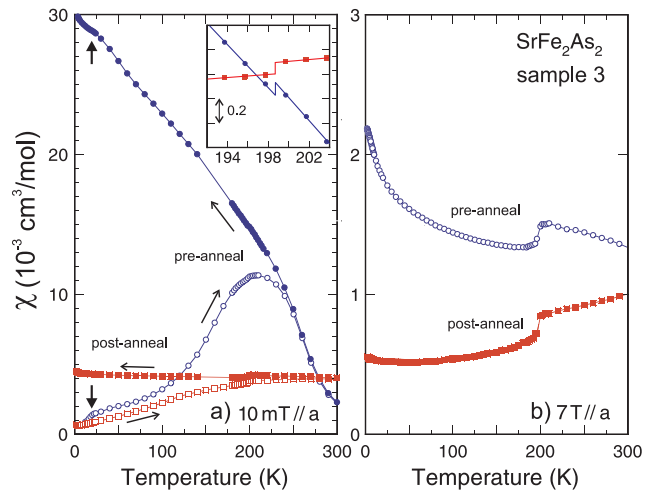}}
  \caption{\label{susc} Magnetic susceptibility of one single-crystal specimen of \Sr\ in 10~mT (a) and 7~T (b) fields applied parallel to the basal plane measured before (blue circles) and after (red squares) 300$^{\circ}$C annealing (open and closed symbols denote zero-field-cooled (ZFC) and field-cooled (FC) conditions, respectively). The SC transition is shown by the arrows. Note the difference in scales of the left and right panels. Inset shows zoom of $\chi(T)$ at structural/magnetic transition before and after annealing (offset for clarity).}
\end{figure}

A systematic study of the magnetic susceptibility of numerous samples provides striking evidence that this stacking fault also has dramatic consequences for the magnetic properties of this material.
Fig.~\ref{susc} presents a representative data set taken using one large single crystal of \Sr\ both before (circles) and after (squares) a 300$^{\circ}$C/2hr annealing treatment, measured in low (a) and high (b) magnetic fields.
We emphasize two striking features present in the as-grown, unannealed sample: a dramatic enhancement of $\chi(T)$ proportional to the superconducting volume fraction, and a large concomitant irreversibility in low-field $\chi(T)$ that persists up to $\sim 250$~K.
After heat treatment of the same sample, these features are both strongly suppressed, yielding a high-field $\chi(T)$ curve (Fig.~\ref{susc}b ``post-anneal'' data) consistent with published results \cite{Yan}, and a low-field curve (Fig.~\ref{susc}a ``post-anneal'' data) with strongly reduced magnitude of irreversibility. This dramatic change in $\chi(T)$ is unlikely originating from sample impurities, whose concentration and magnetic behavior are surely unaffected by such modest annealing temperatures. Moreover, as discussed previously, the heat treatment also wipes out any trace of superconductivity, suggesting these magnetic properties are also tied to the presence of lattice distortion.


\begin{figure}[!t] \centering
  \resizebox{7cm}{!}{
  \includegraphics[width=8cm]{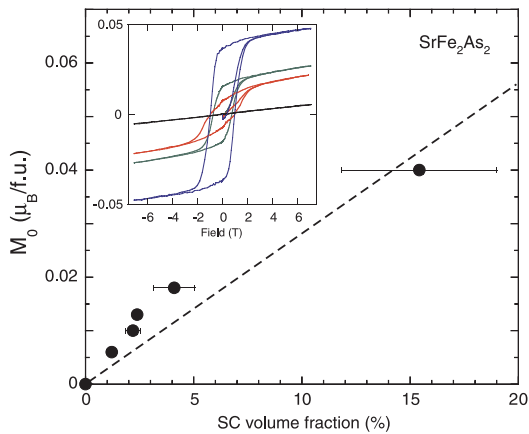}}
  \caption{\label{vf} Ferromagnetic moment and superconducting volume fraction at 1.8~K for several samples of \Sr\, exhibiting a proportional relationship.  Inset: magnetization hysteresis loops at 1.8~K. The non-SC sample exhibits linear paramagnetic field-dependence. The ordered moment in SC samples was estimated by linear extrapolation from high fields, while the SC volume fraction was deduced from the slope of the virgin curve in the Meissner state.}
\end{figure}

Intriguingly, the magnetic enhancement of $\chi(T)$ in superconducting samples is {\it ferromagnetic} in nature: as shown in the inset of Fig.~\ref{vf}, magnetization $M$ isotherms shown for several samples at 1.8~K clearly indicate the hysteretic behavior of $M(H)$ saturating to a linear field-dependence near 3~T. Because the high-field slope of the superconducting samples matches the slope of the non-superconducting sample, it is clear that the bulk magnetic response of \Sr\ is maintained in all samples. This is also evident from the fact that the magnetostructural transition at $T_0$ appears to be impervious to either magnetic history, field strength or heat treatments, as shown in inset of Fig.~\ref{susc}a where the abrupt step in $\chi(T)$ is shown to maintain its magnitude ($\sim 2\times10^{-4}$~cm$^3$/mol) after annealing.

In stark contrast, the apparent ordered moment $M_0$ extracted from fits to $M(H)$ strongly varies with sample. In fact, as shown in the main panel of Fig.~\ref{vf}, $M_0$ scales with the superconducting volume fraction, providing direct evidence that the cause of the enhanced susceptibility and irreversibility in $\chi(T)$ is not only related to the existence of lattice distortions, but is altered in direct proportion to the volume fraction of these distortions.
Interestingly, if this linear correlation is extrapolated to 100\% superconducting volume fraction, the corresponding ordered moment would be $\sim 0.5~\mu_B$/f.u., comparable to the size of the AFM staggered moment \cite{Zhao}.
In \Sr, where intrinsic disorder could affect the orthorhombic AFM domain structure expected to form below $T_0$, it is possible that a ferromagnetic moment proportional to the staggered moment arises at these domain boundaries; the presence of planar crystallographic dislocations, as evidenced by our X-ray analysis, could certainly give rise to such a scenario.

Recent studies of pressure-induced superconductivity in the 122 parent compounds have indicated a quite narrow region of full-volume fraction superconductivity \cite{Alireza}, most evident in the case of CaFe$_2$As$_2$ where hydrostatic (He gas) pressure conditions have revealed the absence of SC over any sizeable range of pressure \cite{Yu}, in contrast to that originally reported \cite{CaFe2As2}.
Given the likelihood of strain-induced superconductivity in \Sr, we suspect that a similar mechanism may be at play in all of these materials, especially in the case of \Sr\ where $T_c =21$~K is also reported at the pressure range of peak volume fraction \cite{Alireza}.
In the transition-metal-doped systems, superconductivity is stabilized at $\sim 20\%$ substitution levels presumably through some form of electronic structure shift via charge doping. However, it is unlikely that significant changes in charge distribution arise from the lattice distortions evident in superconducting \Sr\ crystals. Rather, the same mechanism which gives rise to superconductivity under pressure, where charge doping is presumably not occurring, could account for strain-induced superconductivity in these materials. Regarding this, it would be interesting to check for a ferromagnetic moment associated with the superconductivity observed under applied pressure to compare the nature of strain induced by possible non-hydrostatic conditions which may be the cause of this peculiar phase.
In any case, the interplay of structural disorder and internal strain that stabilizes superconductivity and ferromagnetism in \Sr\ prompt further study.


The authors acknowledge B.~W.~Eichhorn for experimental assistance
and R.~L.~Greene, S.~E.~Sebastian and M.~A.~Tanatar for useful
discussions. N.P.B. acknowledges CNAM support.


\end{document}